\documentclass{PoS}

\newcommand{\R}{\mathbb R}

\newcommand{\eg}{e.g.,\ }
\newcommand{\SigmaP}{\texttt{Sigma}}
\newcommand{\HarmonicSumsP}{\texttt{HarmonicSums}}

\newcounter{mmacnt}
\def\restartmma{\setcounter{mmacnt}{0}}
\restartmma \catcode`|=\active
\def|#1|{\mathrm{#1}}
\catcode`|=12
\newenvironment{mma}{
 \par\smallskip
 \catcode`|=\active
 \parskip=0pt\parindent=0pt % locally
 \small
 \def\In##1\\{%
   \def\linebreak{\hfill\break\null\qquad}%
   \refstepcounter{mmacnt}
   \hangindent=2.5em\hangafter=0
   \leavevmode
   \llap{\tiny\sffamily In[\arabic{mmacnt}]:=\kern.5em}%
   \mathversion{bold}\footnotesize$\displaystyle##1$\normalsize
   \mathversion{normal}\par
 }%
 \def\Print##1\\{%
   \def\linebreak{\hfill\break}%
   \hangindent=2.5em\hangafter=0
   \leavevmode ##1\par}%
 \def\Out##1\\{%
   \def\linebreak{$\hfill\break\null\hfill$}%
   \kern\abovedisplayskip\par
   \hangindent=2.5em\hangafter=0
   \leavevmode
   \llap{\tiny\sffamily Out[\arabic{mmacnt}]=\kern.5em}
   \footnotesize$\displaystyle##1$\normalsize\hfill\null\par
   \kern\belowdisplayskip
 }%
 \def\Warning##1##2\\{%
   \def\linebreak{\hfill\break}%
   \hangindent=2.5em\hangafter=0
   \leavevmode
   {\scriptsize##1 : ##2}\par}%
}{%
 \par\smallskip
}

\usepackage{color}
\usepackage{framed}
\usepackage{url}

\newenvironment{fshaded}{%
\MakeFramed {\FrameRestore}
}
{\endMakeFramed}

\title{The package HarmonicSums: Computer Algebra and Analytic aspects of Nested Sums}

\ShortTitle{Computer Algebra and Analytic aspects of Nested Sums}

\author{\speaker{Jakob Ablinger}%
         \thanks{This work has been supported in part by the Austrian Science Fund (FWF) grant SFB F50 (F5009-N15) and by the EU Network LHCPHENOnetPITN-GA-2010-264564.}\\
        Research Institute for Symbolic Computation (RISC)\\
	Johannes Kepler University Linz, Altenberger Stra\ss e 69, A-4040 Linz, Austria \\
        E-mail: \email{jablinge@risc.jku.at}}

\abstract{This paper summarizes the essential functionality of the computer algebra package \ttfamily HarmonicSums\rmfamily . On the one hand \ttfamily HarmonicSums\rmfamily\ can work with nested sums such as harmonic sums and their generalizations and
on the other hand it can treat iterated integrals of the Poincar\'e and Chen-type, such as harmonic polylogarithms and their generalizations. The interplay of these representations and the analytic aspects are illustrated by concrete examples.}

\FullConference{Loops and Legs in Quantum Field Theory\\
		 27 April 2014 - 02 May 2014\\
		 Weimar, Germany}

\begin{document}

\section{Introduction}
This paper is dedicated to the presentation of the basic features of the computer algebra package \ttfamily HarmonicSums\rmfamily\ which was developed in \cite{Da} and \cite{Diss} and which was afterwards extended and generalized. The package \ttfamily HarmonicSums\rmfamily\ was designed to support 
calculations with special nested objects arising in massive higher order perturbative calculations in renormalizable quantum field theories. On the one hand \ttfamily HarmonicSums\rmfamily\ can work with nested sums such as harmonic sums \cite{Bluemlein1999,Vermaseren1998} and their generalizations (S-sums \cite{Ablinger2013,Moch2002}, cyclotomic harmonic sums \cite{Ablinger2011}, cyclotomic S-sums \cite{Diss}, binomial sums \cite{binom2013,binom2013a,AblingerBinom,Davydychev2003,Weinzierl2004}) and
on the other hand it can treat iterated integrals of the Poincar\'e and Chen-type, such as harmonic polylogarithms \cite{Remiddi2000} and their generalizations (multiple polylogarithms \cite{Ablinger2013}, cyclotomic harmonic polylogarithms \cite{Ablinger2011}).
\ttfamily HarmonicSums\rmfamily\ provides functions to compute (generalizations of) the Mellin-transformation of these iterated integrals which leads to the nested sums and on the other hand inverse Mellin transforms of the nested sums can be computed.
\ttfamily HarmonicSums\rmfamily\ offers commands that rewrite certain types of nested sums into expressions in terms of generalized S-sums and it can be used to derive algebraic and structural relations between the nested sums (compare \cite{Diss,Da,Bluemlein2004,Bluemlein2008,Bluemlein2009,Bluemlein2009a}) as well as relations between the values of the sums at infinity and connected to it the values
of the iterated integrals evaluated at special constants (compare \cite{Diss,Ablinger2011,Ablinger2013,Broadhurst2010}). In addition algorithms to compute series expansions (especially asymptotic expansions) of these nested objects are implemented. The package has already been used extensively, for example during the work on \cite{Used1,Used2,Used3,Used4,Used5}.

\section{The package \HarmonicSumsP}

Note that this section contains a whole Mathematica session that runs throughout the sections. The inputs are given in the way how one has to type them into Mathematica and the outputs are
displayed as Mathematica gives them back.
We start the session by loading the package \HarmonicSumsP:
\begin{mma}
{
\In \textnormal{\bf <\hspace{-0.15cm} < HarmonicSums.m}\\
\fbox{\parbox{12cm}{\footnotesize HarmonicSums by Jakob Ablinger -RISC Linz- Version 1.0 (15/05/04)}}
}
\end{mma}

\subsection*{Definition of the Nested Sums}
In the package \HarmonicSumsP\ harmonic sums, S-sums, cyclotomic harmonic sums and cyclotomic S-sums are denoted by the letter \ttfamily S\rmfamily\ as we can see in the following examples.\\
The command \ttfamily ToHarmonicSumsSum\rmfamily\ yields the definition of the sums.
\begin{mma}
{
\In  \textnormal{\bf \{S[1, 2, 3, 4, n], S[1, 2, 3, \{2, 1/3, 4\}, n]\}//ToHarmonicSumsSum}\\
\Out {\left\{\sum_{\tau_1=1}^n \frac{
\sum_{\tau_2=1}^{\tau_1} \frac{
\sum_{\tau_3=1}^{\tau_2} \frac{
\sum_{\tau_4=1}^{\tau_3} \frac{1}{\tau_4^4}}{\tau_3^3}}{\tau_2^2}}{\tau_1},\sum_{\tau_1=1}^n \frac{2^{\tau_1} 
\sum_{\tau_2=1}^{\tau_1} \frac{3^{-\tau_2} 
\sum_{\tau_3=1}^{\tau_2} \frac{4^{\tau_3}}{\tau_3^3}}{\tau_2^2}}{\tau_1}\right\}}\\
}
{
\In \textnormal{\bf \{S[\{\{3, 2, 1\}, \{4, 1, 2\}, \{2, 0, -2\}\}, n], S[\{\{3, 2, 1\}, \{4, 1, 2\}, \{2, 0, 2\}\}, \{2, 1/3, -4\}, n]\}//ToHarmonicSumsSum}\\
\Out {\left\{\sum_{\tau_1=1}^n \frac{
\sum_{\tau_2=1}^{\tau_1} \frac{
\sum_{\tau_3=1}^{\tau_2} \frac{(-1)^{\tau_3}}{4 \tau_3^2}}{\big(
        1+4 \tau_2\big)^2}}{2+3 \tau_1},\sum_{\tau_1=1}^n \frac{2^{\tau_1} 
\sum_{\tau_2=1}^{\tau_1} \frac{3^{-\tau_2} 
\sum_{\tau_3=1}^{\tau_2} \frac{(-1)^{\tau_3} 4^{-1+\tau_3}}{\tau_3^2}}{\big(
        1+4 \tau_2\big)^2}}{2+3 \tau_1}\right\}}\\
}
\end{mma}
Note that for internal reasons, sometimes the name \ttfamily CS\rmfamily\ is used to denote cyclotomic harmonic sums and cyclotomic S-sums.
\begin{mma}
{
\In \textnormal{\bf \{CS[\{\{3, 2, 1\}, \{4, 1, 2\}, \{2, 0, -2\}\}, n], CS[\{\{3, 2, 1\}, \{4, 1, 2\}, \{2, 0, 2\}\}, \{2, 1/3, -4\}, n]\}//ToHarmonicSumsSum}\\
\Out {\left\{\sum_{\tau_1=1}^n \frac{
\sum_{\tau_2=1}^{\tau_1} \frac{
\sum_{\tau_3=1}^{\tau_2} \frac{(-1)^{\tau_3}}{4 \tau_3^2}}{\big(
        1+4 \tau_2\big)^2}}{2+3 \tau_1},\sum_{\tau_1=1}^n \frac{2^{\tau_1} 
\sum_{\tau_2=1}^{\tau_1} \frac{3^{-\tau_2} 
\sum_{\tau_3=1}^{\tau_2} \frac{(-1)^{\tau_3} 4^{-1+\tau_3}}{\tau_3^2}}{\big(
        1+4 \tau_2\big)^2}}{2+3 \tau_1}\right\}}\\
}
\end{mma}
In addition \HarmonicSumsP\ can deal with binomial sums, which are denoted by \ttfamily BS\rmfamily:
\begin{mma}
{
\In \textnormal{\bf BS[\{\{2, 1, 3\}, \{1, 0, 1\}\}, \{4, 3\}, \{\{\{2\}, \{1, 1\}\}, \{\{1, 1\}, \{2\}\}\}, n]//ToHarmonicSumsSum}\\
\Out {\sum_{\tau_1=1}^n \frac{4^{\tau_1} \big(
        2 \tau_1\big)! 
\sum_{\tau_2=1}^{\tau_1} \frac{3^{\tau_2} \big(
        \tau_2!\big)^2}{\big(
        2 \tau_2\big)! \tau_2}}{\big(
        \tau_1!\big)^2 \big(
        1+2 \tau_1\big)^3}}\\
}
\end{mma}
Hence a summand of the form $\frac{x^{\tau_i}}{(a \tau_i + b)^c}\frac{(f_1 \tau_i)!\cdots (f_j \tau_i)!}{(g_1 \tau_i)!\cdots (g_k \tau_i)!}$ is represented by $\{a,b,c\}$ in the first, $x$ in the second and $\{\{f_1,\ldots,f_j\},\{g_1,\ldots,g_k\}\}$ in the third index set.

\subsection*{Transformation to Nested Sums}
Using the command \ttfamily TransformToSSums\rmfamily\ an extension \cite{Ablinger2014} of the algorithm described in \cite{Diss} is performed to rewrite nested sum expressions in terms of harmonic 
sums, S-sums, cyclotomic harmonic sums and cyclotomic S-sums.
\begin{mma}
{
\In {\textnormal{\bf{$\sum_{i=1}^n \frac{2 (5+2 i) 
\sum_{j=1}^i \frac{1}{j^2}}{\big(
        2+3 i+i^2
\big)
\big(12+7 i+i^2\big)}$}}\textnormal{\bf//TransformToSSums}}\\
\Out {\frac{1}{(n+1) (n+2) (n+3) (n+4)} \frac{1}{54} \big(
        -n \big(
                43 n^3+394 n^2+1163 n+1100\big)
        +36 (n+1)^2 \big(
                n^2+8 n+15\big) S_2(n)
\big)}\\
}
\end{mma}

\subsection*{Definition of the Nested Integrals}
Harmonic polylogarithms, multiple polylogarithms and cyclotomic harmonic polylogarithms are denoted by the letter \ttfamily H\rmfamily\ as we can see in the following examples. The command 
\ttfamily To\-Harmonic\-Sums\-Inte\-grate\rmfamily\ yields the definition of the nested integrals.
\begin{mma}
{
\In \textnormal{\bf \{H[1,2,-3,4,x], H[\{3,1\},\{5,1\},\{2,0\},x]\}//ToHarmonicSumsIntegrate}\\
\Out {\left\{\int_0^\textnormal{x} \frac{\int_0^{\tau_1} \frac{\int_0^{\tau_2}\frac{\int_0^{\tau_3} \frac{1}{\tau_4-4} \, d\tau_4}{\tau_3+3} \, d\tau_3}{\tau_2-2} \, d\tau_2}{\tau_1-1} \, d\tau_1, \int_0^\textnormal{x} \frac{\tau_1 \left(\int_0^{\tau_1} \frac{\tau_2\left(\int_0^{\tau_2} \frac{1}{\tau_3+1} \,d\tau_3\right)}{\tau_2^4+\tau_2^3+\tau_2^2+\tau_2+1} \,d\tau_2\right)}{\tau_1^2+\tau_1+1} \, d\tau_1\right\}}\\
}
\end{mma}
Note that an index $a\in\mathbb R$ yields an iteration over $\frac{1}{\textnormal{sign}(a) \tau_i - a},$ while an index $\{l,k\}$ $l,k\in \mathbb N$ indicates an iteration over $\frac{\tau_i ^k}{\Phi_l(\tau_i)},$ where $\Phi_l$ is the $l-$th cyclotomic polynomial.
For iterations over more general functions the name \ttfamily GL\rmfamily\ is reserved. The functions can then be defined using \ttfamily VarGL:\rmfamily\
\begin{mma}
{
\In \textnormal{\bf GL[\{$\sqrt{1-\textnormal{ \tiny VarGL}}$,$\frac{1}{\textnormal{{\tiny VarGL}}+1}$\},x]//ToHarmonicSumsIntegrate}\\
\Out {\int_0^x \left(\int_0^{\tau _1} \frac{1}{1+\tau _2} \, d\tau _2\right) \sqrt{1-\tau _1} \, d\tau _1}\\
}
\end{mma}

\subsection*{Shuffle and Quasi-Shuffle Product}
The functions \ttfamily LinearExpand\rmfamily\ and \ttfamily LinearHExpand\rmfamily\ are provided to expand products of harmonic sum, S-sums, cyclotomic harmonic sums and cyclotomic S-sums and products 
of harmonic polylogarithms, multiple polylogarithms and cyclotomic harmonic polylogarithms, respectively.
\begin{mma}

{
\In \textnormal{\bf S[\{\{3, 2, 1\}, \{2, 0, -2\}\}, \{1, 4\}, n] S[\{\{3, 1, 1\}\}, \{-3\}, n]//LinearExpand}\\
\Out {-\textnormal{S}[\{\{3, 1, 1\}, \{2, 0, -2\}\}, \{-3, 4\}, n] + 
 \textnormal{S}[\{\{3, 2, 1\}, \{2, 0, -2\}\}, \{-3, 4\}, n] +\\
 \textnormal{S}[\{\{3, 1, 1\}, \{3, 2, 1\}, \{2, 0, -2\}\}, \{-3, 1, 4\}, n] + 
 \textnormal{S}[\{\{3, 2, 1\}, \{2, 0, -2\}, \{3, 1, 1\}\}, \{1, 4, -3\}, n] + \\
 \textnormal{S}[\{\{3, 2, 1\}, \{3, 1, 1\}, \{2, 0, -2\}\}, \{1, -3, 4\}, n]}\\
}
{
\In \textnormal{\bf H[1, 2, x] H[3, 4, x]//LinearHExpand}\\
\Out {\textnormal{H}[1, 2, 3, 4, \textnormal{x}] + \textnormal{H}[1, 3, 2, 4, \textnormal{x}] + \textnormal{H}[1, 3, 4, 2, \textnormal{x}] + 
 \textnormal{H}[3, 1, 2, 4, \textnormal{x}] + \textnormal{H}[3, 1, 4, 2, \textnormal{x}] + \textnormal{H}[3, 4, 1, 2, \textnormal{x}]}\\
}
\end{mma}

\subsection*{Differentiation of Nested Sums}
In order to differentiate expressions involving harmonic sums, S-sums or cyclotomic harmonic sums the Mathematica function \ttfamily D\rmfamily\ is extended. Note that here we actually work with the analytic continuation of these sums; for details see \eg \cite{Diss,Ablinger2013,Bluemlein2009,Bluemlein2009a}.
\begin{mma}
{
\In \textnormal{\bf D[S[3, 1, n] + n S[3, \{2\}, n] - S[\{\{2, 1, 1\}\}, n], n]}\\
\Out {1 + (1 + 2\;n)^{-2} - (3 + 2\;n)^{-2} + (1 + 2\;(1 + n))^{-2} + \textnormal{S}[\{\{2, 1, 2\}\}, n] + \textnormal{H}[\{0, 0\}, \{1, 0\}, 1] - \frac{\textnormal{S}[2, n]}{4}\\ + \textnormal{H}[\{2, 0\}, \{0, 0\}, 1] + \textnormal{S}[2, 2\;n] + \textnormal{S}[2, \infty]\;\textnormal{S}[3, n] - \textnormal{S}[2, \infty]\;\textnormal{S}[3, \infty] + 
 9\;\frac{\textnormal{S}[5, \infty]}{2} - \textnormal{S}[3, 2, n] + \textnormal{S}[3, \{2\}, n] - 3\;\textnormal{S}[4, 1, n] + n\;((\textnormal{H}[0, 0, 1] - \textnormal{H}[0, 0, 2])\;\textnormal{H}[1, 0, 1] + (-\textnormal{H}[0, 0, 1] + \textnormal{H}[0, 0, 2])\;\textnormal{H}[1, 0, 1] - \textnormal{H}[0, 0, 1, 0, 2] + \textnormal{H}[0, 2]\;\textnormal{S}[3, \{2\}, n] - 3\;\textnormal{S}[4, \{2\}, n])}\\
}
\end{mma}

\subsection*{Basis Representations}
For computing basis representations of harmonic sums, S-sums, cyclotomic harmonic sums, harmonic polylogarithms, cyclotomic polylogarithms or multiple polylogarithms \HarmonicSumsP\ provides the functions \ttfamily Com\-pu\-te\-HSum\-Ba\-sis, ComputeSSumBasis, Com\-pute\-CSum\-Ba\-sis\rmfamily\ and 
\ttfamily Com\-pute\-H\-Log\-Ba\-sis\rmfamily\ are provided.
\begin{itemize}
 \item \ttfamily ComputeHSumBasis[w,n]\rmfamily\ computes a basis and the corresponding relations for harmonic sums at weight \ttfamily w\rmfamily. With the options 
    \ttfamily UseDifferentiation\rmfamily\ and \ttfamily Use\-Half\-Inte\-ger\rmfamily\ it can be specified whether relations due to differentiation and argument duplication should be used.
 \item \ttfamily ComputeSSumBasis[w,x,n]\rmfamily\ computes a basis and the corresponding relations for S-sums at weight \ttfamily w\rmfamily\ where the allowed $``x``-$ indices are defined in the list 
    \ttfamily x\rmfamily. With the options \ttfamily UseDifferentiation\rmfamily\ and \ttfamily UseHalfInteger\rmfamily\ it can be specified whether relations due to 
    differentiation and argument duplication should be used.
 \item \ttfamily ComputeCSumBasis[w,{let},n]\rmfamily\ computes a basis and the corresponding relations for cyclotomic harmonic sums at weight \ttfamily w\rmfamily\ with letters \ttfamily let\rmfamily. 
    With the options \ttfamily Use\-Diff\-erent\-ia\-tion, UseMultipleInteger\rmfamily\ and \ttfamily UseHalfInteger\rmfamily\ it can be specified whether relations due to differentiation and 
    argument multiplication should be used.
\item \ttfamily ComputeHLogBasis[w,n]\rmfamily\ computes a basis and the corresponding relations for multiple polylogarithms at weight \ttfamily w\rmfamily. 
    The option \ttfamily Alphabet->a\rmfamily\ and \ttfamily IndexStructure->i\rmfamily\ can be used to specify an alphabet or a special index structure respectively.
\end{itemize}

\begin{mma}
{
\In {\textnormal{\bf ComputeCSumBasis[2, \{\{2, 1\}\}, n, UseDifferentiation -> False,}\\
    \textnormal{\bf UseMultipleInteger -> False, UseHalfInteger -> False}}\\
\Out {\biggl\{
\bigl\{\textnormal{S}[\{\{2,1,-2\}\},n ],\textnormal{S}[\{\{2,1,2\}\},n ],\textnormal{S}[\{\{2,1,-1\},\{2,1,1\}\},n ]\bigr\}, \\
\bigl\{\textnormal{S}[\{\{2,1,1\},\{2,1,1\}\},n ]\to \frac{1}{2} \textnormal{S}[\{\{2,1,1\}\},n ]{}^2+\frac{1}{2} \textnormal{S}[\{\{2,1,2\}\},n ], \\
	\textnormal{S}[\{\{2,1,1\},\{2,1,-1\}\},n ]\to \textnormal{S}[\{\{2,1,-2\}\},n ]+\textnormal{S}[\{\{2,1,-1\}\},n ] \textnormal{S}[\{\{2,1,1\}\},n ]\\-\textnormal{S}[\{\{2,1,-1\},\{2,1,1\}\},n ],
	\textnormal{S}[\{\{2,1,-1\},\{2,1,-1\}\},n ]\to \frac{1}{2} \textnormal{S}[\{\{2,1,-1\}\},n ]{}^2+\\ \frac{1}{2} \textnormal{S}[\{\{2,1,2\}\},n ]
\bigr\}
\biggr\}}\\
}
\end{mma}
In order to look for relations for harmonic sums, S-sums and cyclotomic harmonic sums at infinity we can use the functions \ttfamily ComputeHSumInfBasis, ComputeSSumInfBasis\rmfamily\ and 
\ttfamily ComputeCSumInfBasis\rmfamily\ while for looking for relations between multiple polylogarithms and cyclotomic polylogarithms at $1$ the functions \ttfamily ComputeGeneralizedH1Basis\rmfamily\ and 
\ttfamily Com\-pute\-Cyclo\-tomic\-H1\-Basis\rmfamily\ are provided.
\begin{mma}
{
\In \textnormal{\bf ComputeCSumInfBasis[2, \{\{2, 1\}\}]}\\
\Out {\biggl\{
\bigl\{\textnormal{S}[\{\{2,1,-2\}\},\infty ],\textnormal{S}[\{\{2,1,2\}\},\infty ],\textnormal{S}[\{\{2,1,-1\},\{2,1,1\}\},\infty ]\bigr\}, \\
\bigl\{\textnormal{S}[\{\{2,1,1\},\{2,1,1\}\},\infty ]\to \frac{1}{2} \textnormal{S}[\{\{2,1,1\}\},\infty ]{}^2+\frac{1}{2} \textnormal{S}[\{\{2,1,2\}\},\infty ], \\
	\textnormal{S}[\{\{2,1,1\},\{2,1,-1\}\},\infty ]\to \textnormal{S}[\{\{2,1,-2\}\},\infty ]+\textnormal{S}[\{\{2,1,-1\}\},\infty ] \textnormal{S}[\{\{2,1,1\}\},\infty ]\\-\textnormal{S}[\{\{2,1,-1\},\{2,1,1\}\},\infty ],
	\textnormal{S}[\{\{2,1,-1\},\{2,1,-1\}\},\infty ]\to \frac{1}{2} \textnormal{S}[\{\{2,1,-1\}\},\infty ]{}^2+\\ \frac{1}{2} \textnormal{S}[\{\{2,1,2\}\},\infty ]
\bigr\}
\biggr\}}\\
}
\end{mma}
For harmonic sums and cyclotomic harmonic sums tables with relations are provided \cite{Ablinger2011,Ablinger2014a}. These tables can be applied using the command \ttfamily ReduceToBasis\rmfamily . With the options \ttfamily UseDiff- erentiation\rmfamily\ 
and \ttfamily UseHalfInteger\rmfamily\ it is possible to specify whether relations due to differentiation and argument duplication should be used.

\ttfamily ReduceToBasis[expr,n,Dynamic->True]\rmfamily\ computes relations for harmonic sums, S-sums and cyclotomic harmonic sums in \ttfamily expr\rmfamily\ from scratch 
and applies them while \ttfamily ReduceTo- Basis[expr,n,Dynamic->Automatic]\rmfamily\ uses the precomputed tables and computes relations that exceed the tables form scratch.

\ttfamily ReduceToHBasis\rmfamily\ uses precomputed tables with relations between harmonic polylogarithms and applies them to expressions involving harmonic polylogarithms similar as for \ttfamily Reduce- ToBasis\rmfamily\ the options \ttfamily Dynamic->Automatic/True\rmfamily\ can be set.

\ttfamily ReduceConstants\rmfamily\ uses precomputed tables with relations between harmonic polylogarithms at argument $1$ and harmonic sums at infinity to reduce the appearing 
constants as far as possible again the options \ttfamily Dynamic->Automatic/True\rmfamily\ can be set.

\begin{mma}
{
\In \textnormal{\bf ReduceToBasis[S[2, 1, n] + S[1, 2, n], n]}\\
\Out {\textnormal{S}[1, n] \textnormal{S}[2, n] + \textnormal{S}[3, n]}\\
}
{
\In \textnormal{\bf ReduceToBasis[S[5, 5, \{3, 3\}, n], n, Dynamic -> True]}\\
\Out {\frac{1}{2} \left(\textnormal{S}[5,\{3\},n]^2+\textnormal{S}[10,\{9\},n]\right)}\\
}
{
\In \textnormal{\bf ReduceToHBasis[H[1, 0, x] + H[0, 1, x]]}\\
\Out {\textnormal{H}[0,\textnormal{x}] \textnormal{H}[1,\textnormal{x}]}\\
}
{
\In \textnormal{\bf ReduceConstants[S[1,1,1,1,1,1,1,1, $\infty$] + 2 H[1, 0, 1] + H[0, 1, -1, 1], Dynamic ->  Automatic]}\\
\Out {\frac{5}{201600}\big( \textnormal{S}[1,\infty]^8+140 \textnormal{S}[2,\infty] \textnormal{S}[1,\infty]^6
  +560 \textnormal{S}[3,\infty] \textnormal{S}[1,\infty]^5
  +1890 \textnormal{S}[2,\infty]^2 \textnormal{S}[1,\infty]^4\\
  +1120 (5 \textnormal{S}[2,\infty] \textnormal{S}[3,\infty]+6 \textnormal{S}[5,\infty]) \textnormal{S}[1,\infty]^3
  +20 \left(549 \textnormal{S}[2,\infty]^3+280 \textnormal{S}[3,\infty]^2\right) \textnormal{S}[1,\infty]^2\\
  +720 \left(21 \textnormal{S}[3,\infty] \textnormal{S}[2,\infty]^2+28 \textnormal{S}[5,\infty] \textnormal{S}[2,\infty]+40 \textnormal{S}[7,\infty]\right) \textnormal{S}[1,\infty]
  +7893 \textnormal{S}[2,\infty]^4\\-5600 \textnormal{S}[2,\infty] \left(-\textnormal{S}[3,\infty]^2+54 \textnormal{S}[-1,\infty]+72\right)
  +1680 (8 \textnormal{S}[3,\infty] (\textnormal{S}[5,\infty]-15)+15 \textnormal{S}[8,\infty])\big)}\\
}
\end{mma}

\subsection*{Series Expansions}
The function \ttfamily HarmonicSumsSeries[expr,\{n,p,ord\}]\rmfamily\ can be used to compute series expansions about the point \ttfamily n=p\rmfamily\ of expressions \ttfamily expr\rmfamily\ involving 
harmonic sums, S-sums, cyclotomic harmonic sums, harmonic polylogarithms, multiple polylogarithms and cyclotomic 
harmonic polylogarithms up to a specified order \ttfamily ord.\rmfamily\

\begin{mma}
{
\In \textnormal{\bf HarmonicSumsSeries[n*S[2, n] + n*H[-2, n], \{n, 0, 4\}] // ReduceConstants}\\
\Out {n^4 \left(4\; \textnormal{z5}+\frac{1}{24}\right)+n^3 \left(-\frac{6\; \textnormal{z2}^2}{5}-\frac{1}{8}\right)+n^2 \left(2\;\textnormal{z3}+\frac{1}{2}\right)}\\
}

{
\In \textnormal{\bf HarmonicSumsSeries[n*S[2, n] + n*H[-2, n], \{n, $\infty$, 4\}] // ReduceConstants}\\
\Out {-n\; \textnormal{H}[0,2] +n\; \textnormal{H}[0,n]-\frac{4}{n^3}+\frac{5}{2\; n^2}+n\; \textnormal{z2}-\frac{3}{2 n}+1}\\
}{
\In \textnormal{\bf HarmonicSumsSeries[S[3, 1, \{1/2, 1/3\}, n], \{n,$\infty$,3\}]}\\
\Out {\textnormal{S}[1,\left\{\frac{1}{3}\right\},\infty] \left(-\textnormal{S}[3,\left\{\frac{1}{6}\right\},\infty]+\textnormal{S}[3,\left\{\frac{1}{2}\right\},\infty]
+6^{-n} \left(\frac{1}{5 n^3}-\frac{3^n}{n^3}\right)\right)+\\6^{-n} \left(\frac{12}{25 n^3}-\frac{1}{5 n^2}\right) \textnormal{S}[2,\left\{\frac{1}{3}\right\},\infty]
+6^{-n} \left(\frac{42}{125 n^3}-\frac{6}{25 n^2}+\frac{1}{5 n}\right) \textnormal{S}[3,\left\{\frac{1}{3}\right\},\infty]
+\\ \textnormal{S}[2,\left\{\frac{1}{6}\right\},\infty] \textnormal{S}[2,\left\{\frac{1}{3}\right\},\infty]-\textnormal{S}[1,\left\{\frac{1}{6}\right\},\infty]
 \textnormal{S}[3,\left\{\frac{1}{3}\right\},\infty]-\textnormal{S}[4,\left\{\frac{1}{3}\right\},\infty]+\textnormal{S}[1,3,\left\{\frac{1}{6},2\right\},\infty]
+\\6^{-n} \left(\frac{1}{5 n^2}-\frac{12}{25 n^3}\right) \textnormal{H}[0,3,1]+6^{-n} \left(-\frac{42}{125 n^3}+\frac{6}{25 n^2}-\frac{1}{5 n}\right) \textnormal{H}[0,0,3,1]-\frac{\textnormal{H}[3,1] 6^{-n}}{5
   n^3}}\\
}
\end{mma}
Note that $z2,z3,\ldots$ are used as abbreviations for $S[2,\infty],S[3,\infty],\ldots$ respectively.
In order to compute asymptotic expansions of an S-sums S$[a_1,a_2,\ldots,\{x_1,x_2,\ldots\},n]$ with $|x_i|>1$ for at least one $i$ the option \ttfamily PrincipalValue -> True\rmfamily\ has to be set:
\begin{mma}
{
\In \textnormal{\bf HarmonicSumsSeries[S[1,1,\{2,1\},n],\{n, $\infty$, 3\}]}\\
\Out {\textnormal{S}[1,1,\left\{2,1\right\},n]}\\
}
{
\In \textnormal{\bf HarmonicSumsSeries[S[1,1,\{2,1\},n],\{n, $\infty$, 3\},PrincipalValue->True]}\\
\Out {\frac{1}{2} \left(
        2^n \left(
                -\frac{2}{n^2}-\frac{43}{3 n^3}\right)
        -\frac{\pi ^2}{2}
        +2^n \left(
                \frac{4}{n}+\frac{4}{n^2}+\frac{12}{n^3}\right) 
  \textnormal{LG}[n] 
  \right)}\\
}
\end{mma}
Note that the function $\textnormal{LG}$ is defined as $\textnormal{LG}[n]:=\log(n)+\gamma,$ where $\gamma$ is the Euler-Mascheroni constant.
For computing asymptotic expansions of expressions of the form $\int_0^1x^n\textnormal{GL}[a,x]dx$ for $n\to \infty$ the command \ttfamily GLExpansion\rmfamily\ is provided:
\begin{mma}
{
\In \textnormal{\bf GLExpansion[GL[\{$\sqrt{1+\textnormal{\tiny VarGL}}$\}, x],x,n,ord]}\\
\Out {-\frac{2}{3 n^5}+\frac{4225}{96 \sqrt{2} n^5}+\frac{2}{3 n^4}-\frac{469}{24 \sqrt{2} n^4}-\frac{2}{3 n^3}+\frac{55}{6 \sqrt{2} n^3}+\frac{2}{3 n^2}-\frac{7 \sqrt{2}}{3 n^2}-\frac{2}{3 n}+\frac{4 \sqrt{2}}{3 n}}\\
}
\end{mma}

The function \ttfamily HToS\rmfamily\ can be used to compute the full power series expansions of harmonic polylogarithms, multiple polylogarithms and cyclotomic harmonic polylogarithms about $0$. 
\ttfamily SToH\rmfamily\ is used to perform the reverse direction.
\begin{mma}
{
\In \textnormal{\bf HToS[\{H[-1, 0, -1, x], H[-3, 0, -1/2, x]\}]}\\
\Out {\left\{\sum _{\iota _1=1}^{\infty } \frac{\textnormal{S}[2,\iota _1] (-\textnormal{x})^{\iota _1}}{\iota
   _1}-\sum _{\iota _1=1}^{\infty } \frac{(-\textnormal{x})^{\iota _1}}{\iota _1^3},\sum _{\iota _1=1}^{\infty } \frac{3^{-\iota _1} (-\textnormal{x})^{\iota _1} \textnormal{S}[2,\{6\},\iota
   _1]}{\iota _1}-\sum _{\iota _1=1}^{\infty } \frac{2^{\iota _1} (-\textnormal{x})^{\iota
   _1}}{\iota _1^3}\right\}}\\
}
{
\In \textnormal{\bf SToH[\{$\sum _{\iota _1=1}^{\infty } \frac{(-x)^{\iota _1} \textnormal{S[6,$\iota _1$]}}{\iota _1}$,$\sum _{\iota _1=1}^{\infty } \frac{3^{-\iota _1} (-x)^{\iota _1} \textnormal{S[2,\{6\},$\iota _1$]}}{\iota _1}$\}]}\\
\Out {\left\{\textnormal{H}[-1, 0, -1, \textnormal{x}] - \textnormal{H}[0, 0, -1, \textnormal{x}],\textnormal{H}[-3, 0, -\frac{1}{2}, \textnormal{x}] - \textnormal{H}[0, 0, -\frac{1}{2}, \textnormal{x}]\right\}}\\
}
\end{mma}
For the more general iterated integrals \ttfamily GL\rmfamily\ the command \ttfamily GLToS\rmfamily\ has to be used, note that this command internally relies on the recurrence solver of the package \SigmaP\ \cite{Schneider2007,SchneiderDiss}. 
\begin{mma}
{
\In \textnormal{\bf GLToS[GL[\{$\sqrt{4-\textnormal{\tiny VarGL}} \sqrt{\textnormal{\tiny VarGL}}$\}, x]]}\\
\Out {\sum_{\textnormal{o}_1=2}^{\infty } -\frac{256 x^{\frac{1}{2} \big(-1+2 \textnormal{o}_1\big)} \big(
        \prod_{\iota_1=1}^{\textnormal{o}_1} \frac{-1+2 \iota_1}{8 \iota_1}\big)\big(-1+\textnormal{o}_1\big) \textnormal{o}_1}{\big(-5+2 \textnormal{o}_1\big)\big(-3+2 \textnormal{o}_1\big)\big(-1+2 \textnormal{o}_1\big)^2}}\\
}
\end{mma}

\subsection*{Mellin Transformation and Inverse Mellin Transformation}
To compute the Mellin transform of a possibly weighted harmonic polylogarithm, multiple polylogarithm (with indices in $\R\setminus(-1,1)\cup\{0\}$) and cyclotomic polylogarithm \ttfamily hlog[x]\rmfamily\ we can use the command 
\ttfamily Mellin[hlog[x],x,n]\rmfamily. 
\begin{mma}
{
\In \textnormal{\bf Mellin[H[1, 0, x]/(1 + x)+H[3,x]/(3-x), x, n]}\\
\Out {-2\;(-1)^n\;\textnormal{S}[3, \infty] + (-1)^n\;\textnormal{S}[-2, -1, \infty] + (-1)^n\;\textnormal{S}[-1, -2, \infty] - (-1)^n\;\textnormal{S}[-1, 2, n] + 3^n\;\textnormal{S}[1, n]\;\textnormal{S}[1, \left\{1/3\right\}, \infty] - 3^n\;\textnormal{S}[1, \left\{1/3\right\}, n]\;\textnormal{S}[1, \left\{1/3\right\}, \infty] - 3^n\;\textnormal{S}[2, \left\{1/3\right\}, \infty] + 
 3^n\;\textnormal{S}[1, 1, \left\{1/3, 1\right\}, \infty] - 3^n\;\textnormal{S}[1, 1, \left\{1, 1/3\right\}, n]}\\
}
\end{mma}
For computing the Mellin transform of more general input \ttfamily expr\rmfamily\ the command \ttfamily Gen\-eral\-Me\-llin[expr,x,n]\rmfamily\ is provided. Note that this command internally relies on the recurrence solver of the package \SigmaP.
\begin{mma}
{
\In \textnormal{\bf GeneralMellin[InvMellGen[S[1, 2, \{2, 1/2\}, n] + S[1, n], n, x], n, x]}\\
\Out { \frac{4 (-1)^n \sqrt{2 \pi } n!}{(2 n+1) (2 n+3) (2 n+5) \big( n-\frac{1}{2}\big)!}\;\textnormal{BS}[\{\{1, 0, 0\}\}, \{-(1/4)\}, \{\{\{2\}, \{1, 1\}\}\}, n]\\
	+\frac{2(3 \big(-5+6 \sqrt{2}\big) +2 \big( -8+13 \sqrt{2}\big) n +\big(-4+8 \sqrt{2}\big) n^2)}{3 (n+1) (2 n+3) (2 n+5)}\\
	+\frac{(-1)^n\sqrt{\pi}n!}{(2n+1)(2n+3)(2n+5)\big(n-\frac{1}{2}\big)!}\big(8\sqrt{2}-15\;\textnormal{GL}[\big\{\textnormal{\tiny VarGL}\sqrt{1+\textnormal{\tiny VarGL}}\big\},1]\big)\\
	+\frac{-\big(-1+3^{n+1}\big) (n+1) \textnormal{GL}[\big\{\frac{1}{3-\textnormal{\tiny VarGL}}\big\},1]+3^{n+1} (n+1) S_1\big({\frac{1}{3}},n\big)+1}{(n+1)^2}}\\
}
\end{mma}    
To compute the inverse Mellin transform of a harmonic sum or a S-sum denoted by \ttfamily sum\rmfamily\ we can use the command \ttfamily InvMellin[sum,n,x]\rmfamily. For a definition of S-sum we refer to~\cite{Diss}.  Note that
\ttfamily $\delta_{1-x}$\rmfamily\ denotes the Dirac-$\delta$-distribution $\delta(1-x)\in D'[0,1]$. For cyclotomic 
harmonic sums and S-sums which are not S-sums \ttfamily InvMellin\rmfamily\ yields an integral representations, where  \ttfamily Mellin[a[x,n]]\rmfamily$:=\int_0^1a(x,n)dx$ and 
\ttfamily Mellin[a[x,n],\{x,c,d\}]\rmfamily$:=\int_c^da(x,n)dx.$
\begin{mma}
{
\In \textnormal{\bf InvMellin[S[1, 2, n], n, x]}\\
\Out {\frac{\textnormal{H}[1,0,\textnormal{x}]}{1-\textnormal{x}}}\\
}
{
\In \textnormal{\bf InvMellin[S[1, 2, \{1, 1/3\}, n], n, x]}\\
\Out {\delta_{1-x} \biggl(-\textnormal{S}[1,\left\{1/3\right\},\infty] \textnormal{S}[2,\left\{1/3\right\},\infty]-2
   \textnormal{S}[3,\left\{1/3\right\},\infty]+\textnormal{S}[1,2,\left\{1/3,1\right\},\infty]+\\ \textnormal{S}[2,1,\left\{1/3,1\right\},\infty]\biggr)
+\frac{3^{-n} \textnormal{S}[2,\left\{1/3\right\},\infty]}{3-\textnormal{x}}-\frac{\textnormal{S}[2,\left\{1/3\right\},\infty]}{1-\textnormal{x}}-\frac{3^{-n}\textnormal{H}[3,0,\textnormal{x}]}{\textnormal{x}-3}}\\
}
{
\In \textnormal{\bf InvMellin[S[\{\{3, 1, 2\}\}, n], n, x]}\\
\Out {-\textnormal{Mellin}\left[\textnormal{x}^{3 n} \textnormal{H}[0,\textnormal{x}]\right]-\frac{1}{3} \textnormal{Mellin}\left[\frac{(\textnormal{x}^{3 n}-1) \textnormal{H}[0,\textnormal{x}]}{\textnormal{x}-1}\right]+\frac{1}{3} \biggl(2\;
   \textnormal{Mellin}\left[\frac{(\textnormal{x}^{3 n}-1) \textnormal{H}[0,\textnormal{x}]}{\textnormal{x}^2+\textnormal{x}+1}\right]+\\ \textnormal{Mellin}\left[\frac{\textnormal{x} (\textnormal{x}^{3 n}-1) \textnormal{H}[0,\textnormal{x}]}{\textnormal{x}^2+\textnormal{x}+1}\right]\biggr)-1}\\
}
\end{mma}
In order to compute an integral representation of an expression \ttfamily expr\rmfamily\ containing general S-sums together with harmonic sums and cyclotomic sums the command \ttfamily InvMellGen[expr,n,x]\rmfamily\ can be used:
\begin{mma}
{
\In \textnormal{\bf CollectMellinGen[InvMellGen[S[1, 2, \{2, 1/2\}, n] + S[1, n], n, x], n, x]}\\
\Out {\textnormal{MellinGen}\left[-\frac{\textnormal{H}[2,0,1] \left(x^n-1\right)}{x-1},\{x,1,2\}\right]+\textnormal{MellinGen}\left[\frac{\left(x^n-1\right) \left(1-\textnormal{H}[2,0,x]\right)}{x-1},\{x,0,1\}\right]}\\
}
\end{mma}
Note that here we use the notation \ttfamily MellinGen[a[x,n],\{x,c,d\}]\rmfamily$:=\int_c^da(x,n)dx.$

\section{Conclusion}
In this paper we summarized some features of the computer algebra package \ttfamily HarmonicSums.\rmfamily\ Due to space limitations we had to restrict to the presentation of the main commands, while there are many more commands implemented. For more information we refer to \cite{Diss,Da,Ablinger2013}. 
The package together with several precomputed tables and a more detailed description can be downloaded at \url{http://www.risc.jku.at/research/combinat/software/HarmonicSums}.

\end{document}